\documentstyle[epsfig, twocolumn, 12pt]{esapub}

\newcommand {\asec} {$^{\prime\prime}$~}
\def\mic{$\,\mu\rm m$~}

\begin{document}

%% Do not remove the following six lines:
\setlength{\parindent}{0pt}
\setlength{\parskip}{ 10pt plus 1pt minus 1pt}
\setlength{\hoffset}{-1.5truecm}
\setlength{\textwidth}{ 17.1truecm }
\setlength{\columnsep}{1truecm }
\setlength{\columnseprule}{0pt}
\setlength{\headheight}{12pt}
\setlength{\headsep}{20pt}
\pagestyle{esapubheadings}

%% Title - should be in capitals:
\title{\bf ULTRA-DEEP MID-IR SURVEY OF A LENSING CLUSTER}

%% If the author list spans more than one line then the {\bf (bold
%% font)} command must be inserted for each line
\author{ {\bf B.~Altieri$^1$, L. Metcalfe$^1$} \\
{\bf J.-P. Kneib$^2$}\\
{\bf B. Mc Breen$^3$}\\
\vspace{2mm} \\
$^1$ESA Astrophysics Division, VILSPA, P.O Box 50727, E-28080 Madrid, Spain\\
$^2$Observatoire Midi-Pyr\'ene\'es, Toulouse, France\\
$^3$Physics Department, University College Dublin, Stillorgan Rd., Dublin 4, 
Ireland
}

\maketitle

\begin{abstract}

We present the first results of mid-infrared (MIR) ultra-deep observations 
towards the lensing cluster Abell 2390 using the ISOCAM infrared camera 
on-board ESA's Infrared Space Observatory (ISO) satellite.
They reveal a large number of luminous MIR sources.
Optical and near-infrared (NIR) cross-identification suggests that almost 
all 15$\mu$m sources and about half of the 7$\mu$m are identified with
distant lensed galaxies.
Thanks to the gravitational amplification these sources constitute the
faintest MIR sources detected. We confirm that the number counts derived 
at 15$\mu$m show a clear excess of sources with respect to the predictions
of a no-evolution model.\\

The possible extension of the NGST instrumentation from the near-IR (1-5$\mu$m)
to the thermal infrared, up to 20$\mu$m 
(as suggested by the {\it NGST task group report, October 1997}) 
would permit study of this new population of dust-enshrouded 
AGN/starburst galaxies detected by ISOCAM, up to very high redshifts and with 
vastly improved spatial resolution. The existence of this population demonstrates
that the discrimination of dust contributions, possible in the MIR, must be an 
important consideration in reaching an understanding of the Universe at high 
redshift. Therefore we stress that the access of NGST to the thermal infrared 
would increase tremendously its scientific potential to study the early 
universe. \vspace {5pt} \\

%% Do not remove the previous commands. Your abstract should 
%% end with \vspace {5pt} \\  

%% Please insert your keywords here.
  Key~words: Gravitational lensing; Mid-infrared (MIR); Source counts.

\end{abstract}

\section{INTRODUCTION}

Great progress in the understanding of physical properties of galaxies
has been achieved with MIR and Far-infrared (FIR) observations
using the ISO satellite (Kessler et al. 1996) and especially with the
ISOCAM camera (Cesarsky et al. 1996).
\footnote{Based on observations with ISO, an ESA project with instruments
funded by ESA Member States (especially the PI countries: France,
Germany, the Netherlands and the United Kingdom) with the participation
of ISAS and NASA}

Deep MIR cosmological surveys have been performed with ISOCAM on empty fields 
such as the  Hubble Deep field (Rowan-Robinson et al. 1997, Aussel et al. 1998) 
and the Lockman Hole (Taniguchi et al. 1997).

In parallel the ISO programme [LMETCALF ARCS] 
invested about 30 hours of Science Operations Centre (SOC) Guaranteed 
Time (GT) aimed at searching for a background IR-luminous galaxy 
population with the aid of the gravitational amplification of known cluster 
lenses. That programme (Metcalfe et al. 1998, in preparation), yielded 
relatively deep ISOCAM images through a number of cluster lenses. Initial 
results for the cores of Abell 370 and Abell 2218 were reported in Metcalfe et 
al. (1997) and Altieri et al. (1997), respectively. The success of those
observations, which unveiled the MIR emission of the gravitationally lensed 
arcs and other less distorted distant galaxies, led us to push ISOCAM to 
its ultimate limit. So we performed, in additional ESA SOC GT, very deep 
observations towards the core of one of the most studied and rich lensing 
clusters: Abell 2390 (z=0.23). 
These new observations are reported here and supersede both in area and
deepness the previous ISOCAM observations of this cluster (L\'emonon et al. 1998)

\section{OBSERVATIONS}

One full ISO revolution (16 hours) was allocated for this imaging programme
in 2 colours using the broad, high-responsivity, filter bands around 7\mic
and 15\mic. 
Observations were split and scheduled in 4 consecutive revolutions 
between December 26 and 29, 1997, to take advantage of the best observation 
window near apogee, where the glitch rate is at minimum. 
The key observational parameters were:

\begin{itemize}
\item a high spatial resolution, with the 3\asec pixel-field-of-view (PFOV), 
combined with the 'microscanning' mode to increase spatial resolution to 
the maximum (stepping by fractions of a physical pixel size)
\item a high redundancy (i.e. the number of different detector pixels seeing
each sky pixel), up to 400, in the central part of the field, in order to get 
the best flat-fielding accuracy and glitch rejection (Altieri et al. 1998).
\end{itemize}

A small sky area of 7'$^{2}$ was probed, and only the inner 5'$^{2}$ had full
sensitivity. Other ISOCAM deep cosmological surveys used the 6\asec PFOV: e.g.
the Hubble Deep Field (Rowan-Robinson et al. 1998, Aussel et al. 1997,
Aussel et al. 1998), 
the Lockman Hole (Taniguchi et al. 1998) and the Marano  fields.
These surveys have a poorer spatial resolution, and their absolute astrometry 
determination of MIR source positions was difficult because of the lack of 
obvious NIR/optical counterparts.

\section{DATA REDUCTION \& ANALYSIS}

The data reduction was done partly with the the IDL based (ISO)CAM Interactive
Analysis Analysis (CIA) package (Ott et al. 1997) and partly using the PRETI 
method (Elbaz, this conference, Starck et al. 1997),
based on a Multiresolution Median Transform (MMT) analysis, 
of the pixel time histories, which offers the possibility to separate 
the different constituents of the signals, and to reject signal baseline drifts 
and glitch-induced responsive transients which can obscure weak sources.

The mosaicing was done using the (now classical) drizzling technique,
for the best registration and coaddition of the images.
The final pixel size is 1\asec, but the minimal source separation stays close 
to the original physical pixel size of 3\asec.

\begin{figure}[!ht]
  \begin{center}
    \leavevmode
    \centerline{\epsfig{file=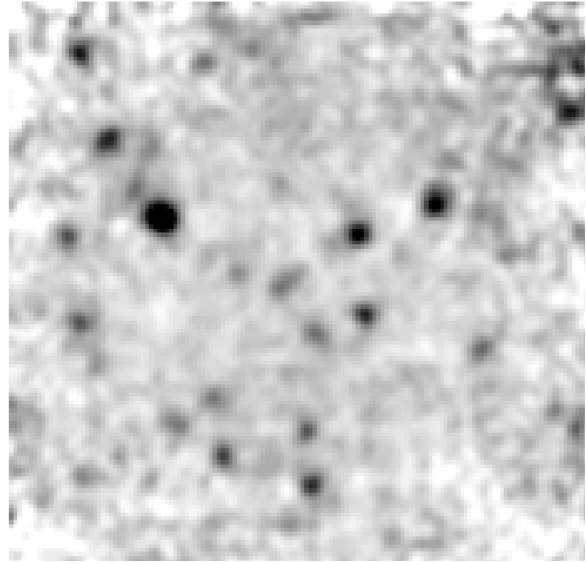,width=8.0cm}}
\vspace{-1.cm}
  \end{center}
  \caption{\em Greyscale plate of the central region of the 7\mic map of Abell 2390}
  \label{lw2_ngst}
\end{figure}

\begin{figure}[!ht]
  \begin{center}
    \leavevmode
    \centerline{\epsfig{file=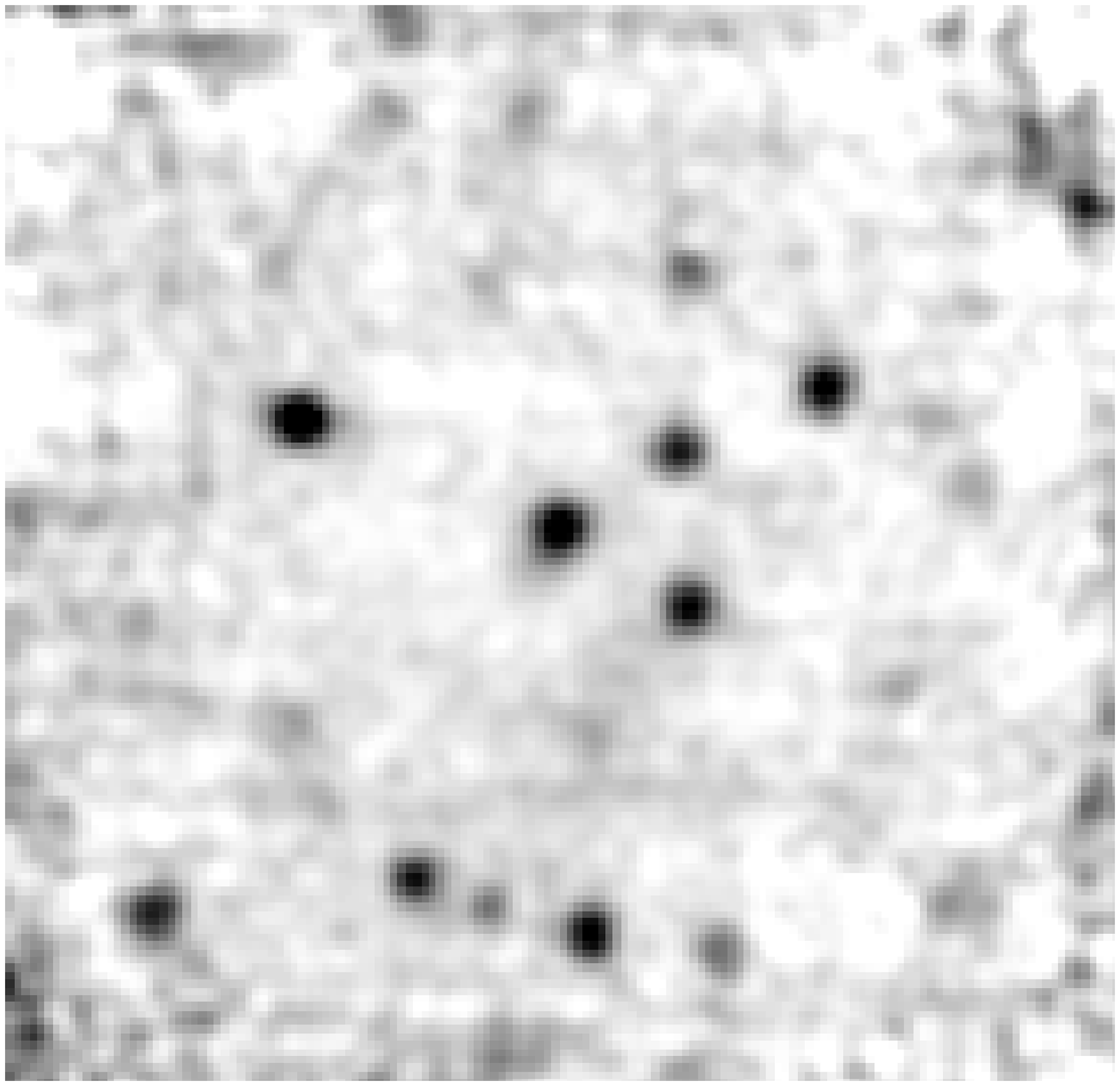,width=8.0cm}}
\vspace{-1.cm}
  \end{center}
  \caption{\em Central region of the 15\mic map of Abell 2390}
  \label{lw3_ngst}
\end{figure}

The central areas of the final maps at 7$\mu$m and 15$\mu$m are shown 
in figures~\ref{lw2_ngst} \& ~\ref{lw3_ngst}. 
The excellent spatial resolution (by ISO standards) allows a positioning 
accuracy to optical-NIR images better than 1\asec, 
by using the 5 brightest sources in the field and unambiguous source
cross-identification with the optical-NIR
The high spatial resolution is the main advantage of this work over
other surveys on `blank' areas of the sky.

%\vspace{8cm}

%\vspace{8cm}

Source catalogues were extracted using the Sextractor package 
(Bertin \& Arnouts 1997), searching for 15 contiguous pixels standing
1.5$\sigma$ above the noise. 
According to our simulations, this scanning of the data returns roughly 
3$\sigma$ sources, because there is some 
correlated noise at short scale due to the 3x3 rebinning of the pixels. 
However on a wide scale the noise in the maps is very close to Gaussian.
The algorithm was also re-run on the negative fluctuations for estimates
of the false-positive detections. Finally simulations of the weakest
detectable faint sources were made by inserting scaled-down versions of 
PSFs into the data, to determine the detection completeness level for 
various source strengths. This was found to be: 80\% at 100$\mu$Jy and 50\% at 
50$\mu$Jy for the 15\mic map. This is close to the limits determined by Aussel
et al. 1998 on the ISOCAM HDF. But thanks to the gravitational magnification, 
these sources constitute the intrinsically faintest MIR sources detected to 
date.

Because of the wide ISOCAM PSF (FWHM = 6\asec at 15\mic) aperture corrected 
photometry was performed on the core of the source PSFs. This method has been 
shown to be linear to the faintest flux levels (Aussel et al. 1998)

% \begin{figure}[!ht]
%  \begin{center}
%    \leavevmode
%    \centerline{\epsfig{file=figure3.ps,width=8.0cm}}
%\vspace{-1.cm}
%  \end{center}
%  \caption{\em Field probed by ISOCAM in several boxes, overlaid on the HST
%  I-band image of the cluster}
%  \label{isocam}
%\end{figure}
%%\vspace{8cm}
%
% \begin{figure}[!ht]
%  \begin{center}
%    \leavevmode
%    \centerline{\epsfig{file=figure4.ps,width=8.0cm}}
%\vspace{-1.cm}
%  \end{center}
%  \caption{\em Equivalent areas corresponding to the boxes of figure 3 at z=1}
%  \label{z1}
%\end{figure}
%%\vspace{8cm}

In order to derive source densities, a detailed lensing model of A2390 has been
produced by Kneib et al. (1998) and was used to determine:
\begin{itemize}
\item the surface-area distortion of the probed area, 
which is a function of the redshift, 
as can be seen in figures~3 and 4
\item the lensing magnification factor, which is also an increasing function 
of redshift has values of 1.5 to 2, but much higher, up to 10, near the 
caustic lines.
\end{itemize}

Therefore it is essential to have a good estimation of the redshifts to
derive the intrinsic fluxes of the sources.
The spectroscopic redshifts were used when available
(P\'ello et al. 1991, Bezecourt \& Soucail 1996),
otherwise the photometric redshifts 
(P\'ello et al. in preparation) and/or 'lensing' predicted redshifts
(Kneib et al. 1998 in preparation)

\section{PRELIMINARY RESULTS}

\subsection{Sources detection}

We derived source catalogues with intrinsic flux densities for the sources, 
detected at 3$\sigma$ confidence level, all with unambiguous optical-NIR 
counterparts on our deep HST-I band image. We found :

\begin{itemize}
\item 32 sources at 7\mic, among which are 19 cluster members and 13 lensed 
sources.
\item 31 sources at 15\mic, that include 3 cluster galaxies and 28 lensed 
sources.
\item 11 sources are detected at both 7\mic and 15\mic. 
\end{itemize}

At 7\mic many E/SO galaxies from the cluster core are detected. This emission 
corresponds to the 5.5\mic restframe emission from the Rayleigh-Jeans tail of 
the cold stellar populations of these elliptical galaxies.

But at 15\mic the core cluster becomes largely transparent and acts as a natural
gravitational telescope, since the 15\mic emission of the elliptical galaxies 
from the cluster core goes below the ISOCAM detection threshold. 
The central cD galaxy is an outstanding exception, 
because it displays very red colors in the MIR,
that could be explained by a very active star forming region located within
the envelope of the galaxy (L\'emonon et al. 1998) or alternatively by
the non-thermal synchrotron emission tail at 15\mic of these radio galaxy.
The absence of 15\mic emission from core cluster galaxies is consistent 
with the observations of A1732 (z=0.193, Pierre et al., 1996), 
where MIR emitters avoid the cluster centre, like galaxies providing the 
blue excess in the Butcher-Oemler effect, and appear to be morphologically 
disturbed or spirals. 
So all but 3 of the 15\mic sources are lensed distant  galaxies.

There is a good agreement between the source counts at 15\mic with ISOCAM 
observations of the HDF down to 50$\mu$Jy, and we can extend the counts down to
30$\mu$Jy when all factors are taken in account (Altieri et al. in 
preparation), as more than 6 sources have intrinsic fluxes below 50 $\mu$Jy, 
in the tiny central area. We derive a preliminary density of sources above 
50$\mu$Jy of 2.5~10$^{4}$ per square degree, 
by taking account of the gravitational effects 
(see \S~3) and the variable depth of the map. These counts lie clearly above 
the no-evolution extrapolation from IRAS counts.

Thanks to the gravitational magnification, these sources constitute the
intrinsically faintest MIR sources detected to date.

\subsection{Lensed Sources Properties}

After the detection of the giant arc at $z=0.724$, and other background objects
in Abell 370 (Metcalfe et al. 1997, and 1998), the ISOCAM capabilities were
confirmed with the nice detections of the complex arc system of Abell 2390
(L\'emonon et al. 1998).
The giant straight arc consists of three parts, {\it A} at $z=1.033$ (Frye \&
Broadhurst, 1998), and {\it B-C}, at $z=0.913$ (Pell\'o et al. 1991).
Source{\it A} was not detected in K band (Smail et al. 1993) 
nor by ISOCAM at either 7\mic nor 15\mic. 
HST images revealed that {\it B} and 
{\it C} are probably two interacting galaxies, and ISOCAM images are in full 
agreement with this picture because the 15\mic emission is centered perfectly 
between the two galaxies with a high [15\mic/7\mic] ratio, 
(L\'emonon et al. 1998).
The source geometry is reminiscent of a distant version of the NGC4038/4039 
interacting system, 
where the most intense starburst in this colliding system
takes place in an off-nucleus region that is inconspicuous at optical 
wavelengths (Mirabel et al. 1998)

Some of the 15\mic sources show very red colours in the NIR 
($R-K > 5$ and $R > 24$) and can be classified as 
``faint red outlier galaxies'' (FROGs), in deep K-band surveys 
(Moustakas et al. 1997).
But other sources can not be
distinguished from the other field galaxies from their optical colours alone,
e.g. the sources do not show any particular blue or red excess. 
This effect was also observed
for the galaxies seen in the Hubble Deep Field detected by ISO (Aussel et 
al., 1997), confirming that star formation is only partially sampled in the 
optical.

At least 3 sources are predicted to be at $z>3$ from their photometric/lensing 
redshift. But it is very interesting to 
note that many other lensed galaxies and arclets that are predicted to be at 
high-z from their optical-NIR photometry were not detected in the MIR. 
Thus ISOCAM 
reveals a different population of objects from those selected in the optical.

\begin{figure}[!ht]
  \begin{center}
    \leavevmode
    \centerline{\epsfig{file=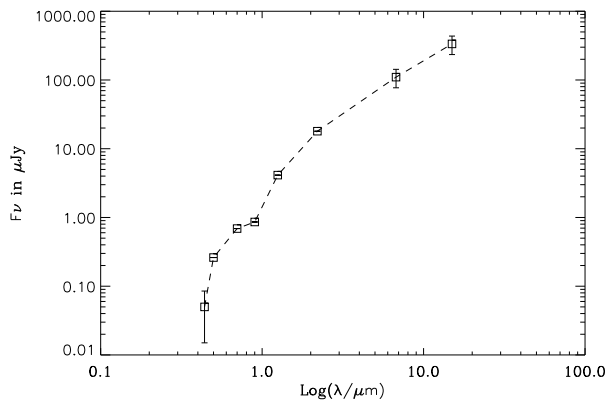,width=8.0cm}}
\vspace{-1.cm}
  \end{center}
  \caption{\em Spectral Energy Distribution of an ISOCAM A2390 source}
  \label{sed260}
\end{figure}

The optical to NIR Spectral Energy Distribution of one of these sources shown 
in figure~\ref{sed260} is strikingly similar to the SED predictions of a 
massive primordial star forming galaxy at an age of 0.2 Gyr (Franceschini et 
al. 1997). Therefore we believe that we have already evidence of a new 
population of MIR emitters at the faintest levels, with very red colours 
([15\mic/7\mic]$>$1).

The dominant morphology of these sources from HST imaging is faint 
disturbed/interacting systems, also with very red optical-NIR colours. 
Some of the sources are too compact for classification. 
It should be noted
that for most of these sources, even the I-band filter probes the 
restframe UV emission, where the morphology can be very different.
There are hints that some of the ISOCAM sources are also detected 
by SCUBA in the sub-mm. 
We could be detecting the same population of dusty star-forming/dusty AGNs 
at intermediate to high redshifts, that show up in the sub-mm (Smail et al.
1997).

\section{CONCLUSION: PROSPECTS FOR THE NGST}      

Our ultra-deep observations with ISOCAM towards the lensing cluster A2390 have
revealed a population of MIR emitters. 
These sources are very faint at optical-NIR wavelengths, 
probably because of dramatic absorption by dust in the restframe UV and optical.
It is very difficult to correlate 
such ISOCAM measurements of hot dust with the 
cooler dust that dominates the energy distribution of the galaxies, 
so the inferred global Star Formation Rate (Rowan-Robinson et al. 1997) 
is very uncertain. But in any case deep 15\mic imaging is a good way to select
star forming regions which are not easily identifiable in UV/optical surveys.
Objects with SED as in figure~\ref{sed260} could not have been found by optical
imaging surveys based on the Lyman-continuum break or on strong emission lines 
(Steidel et al. 1996).

Our observations suggest that star formation in distant galaxies occurs in
different modes and that the most vigorous episodes of star formation probably
arise in dusty environments, as predicted by Franceschini et al. (1994). 
Therefore, great caution must be taken to infer global star formation 
activity based only on continuum luminosity of high-z galaxies, strongly 
absorbed in the UV.
The global star formation history can be fully traced only if the effects of 
dust are taken into account.

The population of 15\mic emitters revealed by ISOCAM would probably show up
with red colours in the 1-5\mic range of the NGST. But their SED is
dramatically increasing through the MIR and towards the FIR and sub-mm, 
where most of their energy is emitted.
So the extension of the NGST instrument capabilities up to the domain of 20\mic 
would tremendously increase the power of the NGST for the observation of these
dusty systems at intermediate and high redshifts. 
With its unprecedented spatial resolution and the access to a wide
wavelength range, the NGST will give the opportunity 
to study the morphology of these galaxies,
to understand the origin of this hot dust emission, 
and to disentangle the scenarios of AGN versus starburst.

A huge scientific potential awaits the NGST with a MIR camera to study the 
early dust-enshrouded universe.
       
\section*{ACKNOWLEDGMENTS}
The ISOCAM data presented in this paper was analyzed using "CIA",     
a joint development by the ESA Astrophysics Division and the ISOCAM
Consortium. The ISOCAM Consortium is led by the ISOCAM PI, C. Cesarsky,
Direction des Sciences de la Matiere, C.E.A., France.

\end{document}